\documentstyle[epsfig,graphicx,12pt,epsf]{article}

\newcommand{\be}{\begin{equation}}
\newcommand{\ee}{\end{equation}}
\newcommand{\ba}{\begin{eqnarray}}
\newcommand{\ea}{\end{eqnarray}}
\newcommand{\ropi}{$\rho \rightarrow \pi^{0} \pi^{0}
\gamma$ }
\newcommand{\roeta}{$\rho \rightarrow \pi^{0} \eta
\gamma$ }
\newcommand{\omepi}{$\omega \rightarrow \pi^{0} \pi^{0}
\gamma$ }
\newcommand{\omeeta}{$\omega \rightarrow \pi^{0} \eta
\gamma$ }

\begin{document}

\begin{center}
{\Large{\bf Chiral loops and VMD in the $V\to PP\gamma$ decays
  }}

\vspace{0.3cm}

\end{center}

\vspace{1cm}

\begin{center}
{\large{J. E. Palomar$^{1}$, S. Hirenzaki$^{1,2}$ and E. Oset$^{1}$ }}
\end{center}

\begin{center}
{\small{$^{1}$ \it Departamento de F\'{\i}sica Te\'orica and IFIC, \\
Centro Mixto Universidad de Valencia-CSIC, \\
Ap. Correos 22085, E-46071 Valencia, Spain}}

{\small{$^{2}$ \it Department of Physics, Nara Women's University, Nara
630-8506, Japan}}

\end{center}

\vspace{1cm}

\begin{abstract}
We evaluate radiative decays of $\rho$ and $\omega$ going to two neutral mesons,
$\pi^{0} \pi^{0}$ and $\pi^{0} \eta$. We use the sequential vector decay
mechanisms in addition to chiral loops and $\rho$-$\omega$ mixing. The chiral
loops are obtained using elements of $U\chi PT$ successfully applied in the study of
meson-meson interactions up to 1.2 GeV. The chiral loops are found very
important in the case of the \ropi decay and  small in the other cases. A good
agreement with present measurements of \ropi and \omepi is obtained and
predictions are made for the other decays where the rates obtained are rather
small.

\end{abstract}

\section{Introduction}

   The radiative vector meson decay into two pseudoscalar mesons has attracted
continuous attention. It has been a case for tests of vector meson dominance
(VMD), through the sequential mechanism $V \to PV \to PP\gamma$
 \cite{singer,Bramon:1992kr}, but more recently it has been advocated as a source
of information on the meson scalar sector in order to learn about the
controversial nature of these states. One of the clearests examples is the
$\phi\to \pi^0\pi^0 \gamma$ decay where the experiment
\cite{Aulchenko:1998xy,Akhmetshin:1999di,Achasov:2000ym} shows very clearly a peak for the
$f_0(980)$
excitation. Similarly the $\phi\to \pi^0\eta \gamma$ reaction shows a clear
peak
for the $a_0(980)$ excitation \cite{Achasov:1998cc}. These decays of
the $\phi$ are particularly interesting since the contribution of the
sequential
processes in the  VMD
model \cite{Bramon:1992kr} is negligible  and the decay width
can be attributed to the excitation of the scalar mesons. Attempts to obtain
the
rates for these reactions and for $\phi\to K^0\bar{K^0} \gamma$ were done
including loops of charged kaons to which the photons could couple
\cite{LucioMartinez:1990uw,Close:1993ay}. A link  to chiral perturbation theory
($\chi PT$) was established in \cite{Bramon:1992ki}, by using the lowest order
chiral meson-meson scattering amplitude. An important step in this direction
was given in
\cite{Marco:1999df} where elements of unitarized chiral perturbation theory
($U
\chi P T$) were used, which directly lead to the excitation of the
$f_0(980)$ and
$a_0(980)$  resonances in these reactions from the consideration of the chiral
loops in coupled channels. The excitation of these resonances from the chiral
loops was made possible because previously it had been found \cite{Oller:1997ti}
that the loop iteration provided by the Bethe-Salpeter equation using a
kernel (potential) from the lowest order chiral Lagrangian, with an appropriate
loop regularization, provides a very good description of the meson-meson
interaction in the scalar sector, including the dynamical generation of the
$\sigma(500)$, $f_0(980)$ and $a_0(980)$  resonances.

The Bethe-Salpeter
approach along the lines of \cite{Oller:1997ti} has been further used to study
meson-meson and meson-baryon interaction in 
\cite{Nieves:2000bx,Nieves:2001wt}. More
elaborate steps taking into account explicitly the contribution of higher order
chiral Lagrangians \cite{Gasser:1985ux}, or generating them from the explicit
exchange of meson resonances,
were done in \cite{OllOsePel,Oller:1999zr} respectively. At the same time these
works provided an explanation of why the Bethe-Salpeter equation with the
explicit use of only the lowest order chiral Lagrangian in \cite{Oller:1997ti} is so
successful in the scalar sector. The interesting thing in all these works is
that
the scalar resonances are generated dynamically from the multiple scattering
implicit in the unitary approach without the need to introduce them as genuine
resonances (those which would survive in the large $N_c$ limit). The nature of
the $\sigma(500)$
as a meson-meson scattering resonance was already advocated in
\cite{Gasser:1991bv,Meissner:1991kz}. Other approaches would start from quark components for the
mesons but the unitarization dresses them with a large cloud of meson-meson
components \cite{torn}. The unitarization of chiral perturbation theory has thus
brought a new insight into the nature of the scalar resonances, which continues
to be a subject of strong debate \cite{eef,kyoto}.

  The purpose of the present paper is to complement the study initiated in
\cite{Marco:1999df} and extend it to other vector meson decays. Here we study
the decays $\rho \to \pi^0\pi^0 \gamma$, $\omega \to \pi^0\pi^0 \gamma$,
$\rho \to \pi^0\eta \gamma$ and  $\omega \to \pi^0\eta \gamma$. 
All these decays were calculated also in ref.~\cite{Prades:1993ys} using
chiral Lagrangians and the extended Nambu-Jona-Lasinio model. The study of the
first two reactions has been done recently in \cite{Bramon:2001un} combining
the
sequential vector meson decay mechanisms with loop contributions. The second of
the reactions has also been revisited recently including effects of
$\rho$-$\omega$ mixing in \cite{Guetta:2001ra}. The latest two reactions were
studied in \cite{Bramon:1992kr} using again the sequential vector meson decay
mechanisms.  
Here we also include the loop corrections and $\rho$-$\omega$ mixing
in the case of the $\rho \to \pi^0\eta \gamma$ reaction. The first of the
reactions has also been studied in \cite{Gokalp:2000ir,Gokalp:2000xy} assuming $\rho \to
\sigma \gamma$ decay, although apparently a too large transition amplitude was
used in the approach, see \cite{Bramon:2001un} for details.

   The present work is also stimulated by the recent measurement of the $\rho
 \to  \pi^0 \pi^0 \gamma$ decay by the SND Collaboration \cite{Achasov:2000zr}
 where a value for the branching ratio

\begin{center}
\begin{equation}
\label{SND}
B(\rho\rightarrow\pi^0\pi^0\gamma)=(4.8^{+3.4}_{-1.8}\pm 0.2)\times 10^{-5}\
\end{equation}
\end{center}

\noindent is obtained. The work of  \cite{Bramon:2001un} finds that there are
 approximately equal
contributions to the  $\rho \to  \pi^0 \pi^0 \gamma$ process from the
sequential
mechanism and from the pion loops.  This latter contribution is very
interesting
since, given the fact that the pion-pion scalar isoscalar amplitude factorizes
on shell in this mechanism, the process is sensitive to the $\pi \pi$
interaction in the region where the sigma meson is produced. Furthermore, the
phase space for the process and the dynamical factors in the total amplitude
make the information appearing there a complement to the one obtained from
other processes from where the $\pi \pi$ phase shifts are measured. Also, the
interference between the loop contribution and the sequential process adds new
information about the pion-pion interaction and the properties of the $\sigma$
meson. In ref. \cite{Bramon:2001un} a simple analytical model for the $\pi \pi$
interaction was used in which  $\sigma(500)$ and $f_0(980)$  exchange  are
explicitly included in the  $\pi \pi$ amplitude. Then values of the
$\sigma(500)$ mass and width from the recent experimental determination
\cite{Aitala:2001xu} are used. These values are
$m_\sigma=478^{+24}_{-23}\pm 17$ MeV and
$\Gamma_{\sigma}=324^{+42}_{-40}\pm 21$. In addition the paper shows that the
results are quite sensitive to the mass and width of the $\sigma$, concluding
that precise measurements of the process can provide valuable information on
these two magnitudes.

   Our aim here is different. We would like to use the reaction as a
further test of the $U \chi PT $ approach to the meson-meson interaction.
Indeed,
analogously to the way the scalar resonances are generated in that approach, in
the present case the reaction mechanisms immediately lead to the excitation of
these resonances without the need to include them explicitly in the formalism.
The theoretical framework allows one to make predictions for production
processes, once the basic parameters of the theory, just one regularizing cut
off in \cite{Oller:1997ti}, have been fixed by fits to the scattering data.

   For the sequential mechanism we follow the approach of
\cite{Bramon:1992kr},
but for the loop contributions we follow the approach of \cite{Marco:1999df}
where the chiral tensor formalism for the vector mesons \cite{Ecker} is used,
as done in
\cite{Huber} in the study of the $\rho \to \pi^+ \pi^- \gamma$ decay. In
\cite{Marco:1999df} it was found
 that the results depend both on the $G_V$ and $F_V$ coupling constants
which are a bit different than the values provided by VMD where $F_V=2G_V$.
 Actually, as shown in \cite{Oset:1999cq}, the
use of the empirical values of $G_V$ and $F_V$ extracted from
$\rho \to \pi \pi$ and $\rho \to e^+
e^-$ decays or from $\phi \to \pi \pi$ and $\phi \to e^+e^-$ decays leads to
appreciable differences in the results in the case of the $\phi \to \pi^0 \pi^0
\gamma$ decay.

\section{Chiral loop contributions}

The chiral loop contribution to the $\rho \rightarrow \pi^0 \pi^0
\gamma$ decay was already formulated in ref. \cite{Marco:1999df}, where the
chiral unitary approach was used to deal with the final state interaction
of the meson-meson system.  We follow the same procedure here and show the
explicit form of the transition amplitudes for all the decays considered
in this paper, $\rho \rightarrow \pi^0 \pi^0 \gamma$,
$\rho \rightarrow \pi^0 \eta \gamma$,
$\omega \rightarrow \pi^0 \pi^0 \gamma$ and
$\omega \rightarrow \pi^0 \eta \gamma$.

We shall make use of the chiral Lagrangian for vector mesons of ref.
\cite{Ecker}
and follow the lines of ref. \cite{Huber} in the treatment of the
radiative meson decay.  The Lagrangian coupling vector mesons to the
pseudoscalar mesons and photons is given by ref. \cite{Ecker},

\begin{center}
\begin{equation}
 {\cal L} = \frac{F_V}{2 \sqrt{2}} < V_{\mu\nu} f^{\mu\nu}_{+} >
+ \frac{iG_V}{\sqrt{2}} < V_{\mu\nu} u^{\mu}u^{\nu} > ,
\label{resolagr}
\end{equation}
\end{center}

\noindent
where $V_{\mu\nu}$ is a 3 $\times$ 3 matrix of antisymmetric tensor fields
representing the octet of vector mesons, $f^{\mu\nu}_{+}$ is related to
the photon field, $u^{\mu}$ are SU(3) matrix involving derivatives of
the pseudoscalar meson fields and $<$ $>$ denotes the trace in flavour space.
The couplings $G_V$ and $F_V$ are deduced from the $\rho \rightarrow
\pi^+ \pi^-$  and $\rho \rightarrow e^+ e^-$ decays, and taken to be
$G_V=69$ MeV and $F_V=154$ MeV \cite{Huber}.  A singlet field, $\omega_1$, is
introduced through the substitution $V_{\mu\nu} \rightarrow V_{\mu\nu} +
I_3 \times \frac{\omega_{1,\mu\nu}}{\sqrt{3}}$ with $I_3$ the 3 $\times$
3 unit matrix, and the physical $\phi$ and $\omega$ meson fields are defined
by
assuming the ideal mixing.
Here the $F_V$ term include the $VPP\gamma$ (V vector and P
pseudoscalar) couplings and the $G_V$ term include $VPP$ and $VPP\gamma$
couplings.

From the Lagrangian, the basic couplings to evaluate the loop
contributions to the present decays
are,

\begin{center}
\begin{eqnarray}
	 t_{\omega K^+K^-}& = & - \frac{1}{2} \frac{G_V M_\omega}{f_\pi^2}
	 (p-p^\prime)_\mu \epsilon^\mu (\omega)
	\nonumber \\
	t_{\omega K^+K^-\gamma} & = & e \frac{G_V M_\omega}{f_\pi^2}
	\epsilon_\nu (\omega)\epsilon^\nu (\gamma)
	\nonumber \\
	& + & \frac{e}{M_\omega f_\pi^2} \left( \frac{F_V}{2} - G_V \right)
	P_\mu  \epsilon_\nu (\omega) [k^\mu \epsilon^\nu (\gamma) - k^\nu
	\epsilon^\mu (\gamma)]
\label{eqn:coupl}
\end{eqnarray}
\end{center}

\noindent
with $p_{\mu}$, $p_{\mu}^{\prime}$ are $K^+$, $K^-$ momenta, $P_{\mu}$ the
$\omega$ meson momentum and $k_{\mu}$ the photon
momentum.
 There are no couplings $\omega \pi^+\pi^-$ nor $\omega
 \pi^+\pi^-\gamma$.  The couplings for $\rho$ meson are given explicitly
 in ref. \cite{Marco:1999df}.
 The pion decay constant $f_{\pi}$ is taken to be 92.4
 MeV.

\begin{figure}
\centerline{\protect\hbox{
\psfig{file=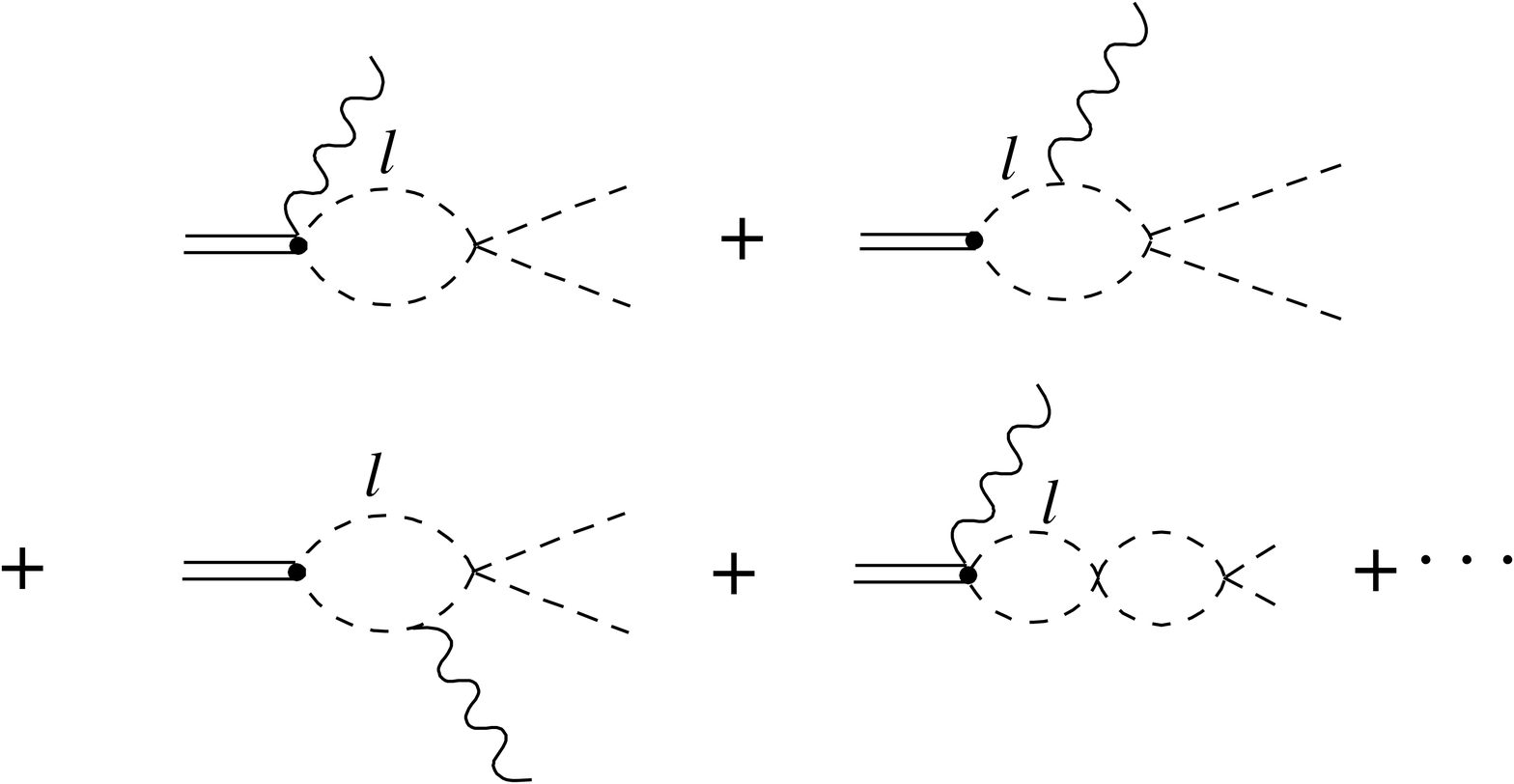,width=0.7\textwidth,silent=}}}
\caption{Loop diagrams included in the chiral loop contributions. The
intermediate states in the loops can be $K^+K^-$ or $\pi^+\pi^-$. }
\label{looplot}
\end{figure}

The $\rho^0$ and $\omega$ decays into neutral mesons and photon can
take place through the loop contributions shown in figure \ref{looplot}.
The technology to introduce the final state interaction is developed in
ref. \cite{Oller98,Marco99} and applied to the radiative decay of
$\rho^0$ and $\phi$  mesons in ref. \cite{Marco:1999df}.
Using gauge invariance arguments one finds that the loop function is
finite and that the meson-meson scattering amplitude factorizes in the
loop integral with its on shell value.
Following the procedure
in ref.
\cite{Marco:1999df}, we can write the explicit expression for the transition
amplitudes for $\rho \rightarrow \pi^0 \pi^0 \gamma$ and $\omega
\rightarrow \pi^0 \pi^0 \gamma$ as,

\begin{center}
\begin{eqnarray}
t_{\rho \rightarrow \pi^0 \pi^0 \gamma}^{loop} & = &
 2e \frac{G_V M_\rho}{f_\pi^2}
\left( \tilde{G}_{\pi\pi} t_{\pi^+\pi^- , \pi^0 \pi^0}
+ \frac{1}{2} \tilde{G}_{KK} t_{K^+K^- , \pi^0 \pi^0}
\right)\epsilon_\mu (\rho)\epsilon^\mu (\gamma)
\nonumber \\
& + &\frac{2e}{f_\pi^2} \left( \frac{F_V}{2} - G_V \right) q
\left( G_{\pi\pi}  t_{\pi^+\pi^- , \pi^0 \pi^0}
+ \frac{1}{2} G_{KK}  t_{K^+K^- , \pi^0 \pi^0}
\right)\epsilon_\mu (\rho)\epsilon^\mu (\gamma)
\nonumber\\
t_{\omega \rightarrow \pi^0 \pi^0 \gamma}^{loop} & = &
 2e \frac{G_V M_\omega}{f_\pi^2}
 \frac{1}{2} \tilde{G}_{KK} t_{K^+K^- , \pi^0 \pi^0}
\epsilon_\mu (\omega)\epsilon^\mu (\gamma)
\nonumber \\
& + &\frac{2e}{f_\pi^2}  \left( \frac{F_V}{2} - G_V \right) q
 \frac{1}{2} G_{KK}  t_{K^+K^- , \pi^0 \pi^0}
\epsilon_\mu (\omega)\epsilon^\mu (\gamma)
\end{eqnarray}
\end{center}

\noindent
where $q$ is the photon momentum in the vector meson rest frame, $\tilde{G}$
the
loop function with the photon attached, $G$ the ordinary two meson
propagator function,  and $t_{M_1 M_2,M_1^\prime M_2^\prime}$
is the strong transition matrix element.  The $\tilde{G}$ is defined in
ref. \cite{Marco:1999df} and has the analytic expression given in
\cite{Oller98}.
The two meson propagator function, $G$ is the one appearing in the
Bethe-Salpeter equation for the meson-meson scattering and is
regularized in ref. \cite{Oller:1997ti} using a cut-off parameter.
The strong transition matrix element, $t_{M_1 M_2,M_1^\prime M_2^\prime}$,
is also evaluated in ref. \cite{Oller:1997ti}.

The transition amplitudes to the $\pi^0\eta\gamma$ final state are readily
obtained by omitting the pion loop contribution for $\rho$ meson and
replacing the $t_{K^+K^- , \pi^0 \pi^0}$ into $t_{K^+K^- , \pi^0 \eta}$ for
both $\rho$ and $\omega$ mesons.

\section{Sequential mechanism contributions}

Apart from the loop contributions studied in section 2, there is another decay
mechanism based on vector meson exchange whose contribution is as important as
the one coming from the loops in the case of \ropi, and dominant in the rest of
the decays studied here. This mechanism has been studied in \cite{Bramon:1992kr,
Bramon:1992ki,Bramon:2001un} and in this section we will follow closely these
references.

\begin{figure} 
\centerline{\protect\hbox{
\psfig{file=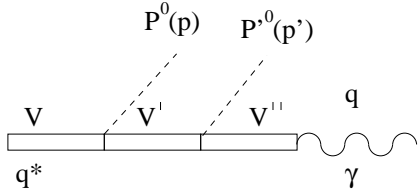,width=0.4\textwidth,silent=}}}
\caption{Feynman diagram corresponding to the vector meson exchange
contribution. V and V' are either $\rho$ or $\omega$ mesons, and P, P' can be
either $\pi^{0}$ or $\eta$ mesons.} 
\label{feyn}
\end{figure}

The Feynman diagrams corresponding to this mechanism, $V(q^{*},\epsilon^{*})
\rightarrow P(p) P'(p') \gamma(q,\epsilon)$, are of the form shown in
figure \ref{feyn}. The vertices involved in these diagrams come from the
Lagrangians:

\begin{center}
\ba
{\cal L}_{VVP} &=& \frac{G}{\sqrt{2}}\epsilon^{\mu \nu \alpha \beta}\langle
\partial_{\mu} V_{\nu} \partial_{\alpha} V_{\beta} P \rangle \nonumber \\
{\cal L}_{V \gamma} &=& -4f^{2}egA_{\mu}\langle QV^{\mu}\rangle
\label{lagr}
\ea
\end{center}

\noindent where $G=\frac{3g^{2}}{4\pi^{2}f}$ is the $\rho \omega \pi$ coupling
and $g$ can be related to the  $G_{V}$ coupling of the chiral resonance
Lagrangians of eq. (\ref{resolagr}), $g=-\frac{G_{V}M_{\rho}}{\sqrt{2}f^{2}}$.
The total amplitude has the form:

\begin{center}
\be
{\cal A}(V \rightarrow P^{0}P'^{0}\gamma)= C_{VPP'\gamma}\left(
\frac{G^{2}e}{\sqrt{2}g}\right) \left\{
\frac{P^{2}\{a\}+\{b(P)\}}{M_{V_1}^{2}-P^{2}-iM_{V_1}\Gamma_{V_1}} 
  + \frac{P'^{2}\{a\}+\{b(P')\}}{M_{V_2}^{2}-P'^{2}-iM_{V_2}\Gamma_{V_2}}
 \right\}
\label{amplitude}
\end{equation}
\end{center}

\noindent where the $\{a\}$ and $\{b\}$ functions and the $C_{VPP'\gamma}$
coefficients are defined in ref. 
\cite{Bramon:1992kr} and $P$ and $P'$ are the momenta of the intermediate
resonance in the $t-$ and $u-$ ($V_1$ and $V_2$) 
channels respectively. The amplitude ${\cal A}$ corresponds to our $t$ matrices
changing the sign.

From the amplitude of eq. (\ref{amplitude}) one can calculate the branching ratios
of the different decay modes. The results are given in table \ref{res}. In the
calculation of the $\omega$ decay channels we have used a momentum-dependent
width for the intermediate $\rho$ meson in the propagators, which leads to
an enhancement of the decay width of around $12\%$ when compared to the
calculation performed with a constant width, as pointed out in refs.
\cite{Bramon:2001un,Guetta:2001ra}.

\subsection{$\rho$-$\omega$ mixing effects}

In addition to the VMD contribution, the incorporation of isospin violation
effects (due to quark mass differences and electromagnetic corrections) allowing
the mixing of the $\rho$ and $\omega$ resonances is readily possible. This
mixing is well known and it has been seen to be relevant in processes like the
$\omega \rightarrow \pi^{+} \pi^{-}$ decay or in the pion form factor in the
$\omega$ region 
\cite{Klingl:1996by,guerrero,Oller:2001ug}. A treatment of this mixing within $\chi PT$ was done
in \cite{Urech:1995ry}. Here we will follow the study of its effects in the radiative
decays of vector resonances done in \cite{Bramon:2001un,Guetta:2001ra}, where the
mixing allows the transition $V \rightarrow V'$ in the process: 
$V \rightarrow V' \rightarrow PP' \gamma$. Thus, the amplitudes can be written as ${\cal A}_{0}(V \rightarrow PP' \gamma) + \epsilon
{\cal A}(V' \rightarrow PP' \gamma)$, where ${\cal A}_{0}(V \rightarrow PP'
\gamma)$ includes the contributions coming from VMD and the loops and  $\epsilon$ is
the mixing parameter:

\begin{center}
\be
\epsilon \equiv
\frac{\Theta^{2}_{VV'}}{M_{V}^{2}-M_{V'}^{2}-i(M_{V}\Gamma_{V}-M_{V'}
\Gamma_{V'})}
\label{mixing}
\ee
\end{center}

\noindent with \cite{Urech:1995ry}

\begin{center}
\be
\Theta_{VV'}^{2}=\frac{M_{V}^{2}}{M_{V'}^{2}}\left[-(m_{K^{0}}^{2}-m_{K^{+}}^{2})
+(m_{\pi^{0}}^{2}-m_{\pi^{+}}^{2})+\frac{e^{2}F_{V}^{2}}{3}\right]
\label{mixparam}
\ee
\end{center}

We have considered this effect in the calculation of the $\omega \rightarrow
\pi^{0} \pi^{0} \gamma$,  \omeeta and
\roeta  decays. There is still another effect of this mixing, which is that it
modifies the $V'$ propagator in ${\cal A}_{0}$ by:

\begin{center}
\be
\frac{1}{D_{V'}(s)} \rightarrow \frac{1}{D_{V'}(s)}\left(
1+\frac{g_{V\pi \gamma}}{g_{V'\pi \gamma}}\frac{\Theta_{VV'}^{2}}{D_{V}(s)}\right)
\label{prop}
\ee
\end{center}

\noindent where $D_{V}=s-M_{V}^{2}+iM_{V}\Gamma_{V}$. This effect is relevant in the
case of the $\rho$ propagator since $g_{\omega\pi \gamma}/g_{\rho\pi \gamma}=3$,
 according to $SU(3)$ symmetry, and in fact makes the branching ratio a $8\%$ larger
in the \omepi  case and a $11\%$ larger in the \omeeta  case.

\section{Numerical results}

Using the transition amplitudes described in the previous sections, we
can calculate the differential decay widths of the $\rho$ and $\omega$
mesons as,

\begin{center}
\begin{equation}
\frac{d\Gamma}{dM_I}= \frac{1}{64 \pi^3} \int_{m_\pi}^{M_V-q-m^\prime}
d\omega \frac{M_I}{M_V^2} \bar{\sum} \sum | t |^2 \theta(1-A^2),
\label{dismi}
\end{equation}
\end{center}

\noindent
where $M_I$ is the invariant mass of the final two mesons, $M_V$ the
initial vector meson mass ($M_{\rho}$ or $M_{\omega}$), $m^\prime$ is the
pion mass for the $\pi\pi\gamma$ decay and $\eta$ mass for the
$\pi\eta\gamma$ decay and $q$ is the photon momentum in the initial
vector meson rest frame.  $A$ accounts for the cosine of the angle between 
the $\pi^0$ and the photon and it is defined as,

\begin{center}
\begin{equation}
A  =\frac{1}{2pq} \left[ (M_V-\omega(p)-q)^2 - m^{\prime 2} -
p^2 - q^2 \right],
\end{equation}
\end{center}

\noindent
where $p$ and $\omega(p)$ are the $\pi^0$ momentum and energy  in the initial
 vector meson rest frame. A
symmetry factor 1/2 must be implemented in eq. (\ref{dismi}) in the case of
$\pi^{0} \pi^{0}$ in the final state.

The spin sum and average of the transition amplitudes, $\bar{\sum} \sum
| t |^2$, can be expressed using the contravariant tensor $F^{ij}$ as:

\begin{center}
\begin{equation}
\bar{\sum} \sum | t |^2 = \frac{1}{3} \left[
F^{ij} F^{ij*} - \frac{1}{|q|^2} (F^{ij}q_j) (F^{i j^\prime *}
q_{j^\prime})
\right],
\end{equation}
\end{center}

\noindent
where the tensor expression $F^{ij}$ of the transition amplitude $t$ is
defined as

\begin{center}
\begin{equation}
t \equiv F^{ij} \epsilon_i (V) \epsilon_j (\gamma).
\end{equation}
\end{center}

\begin{figure} 
\centerline{\protect\hbox{
\psfig{file=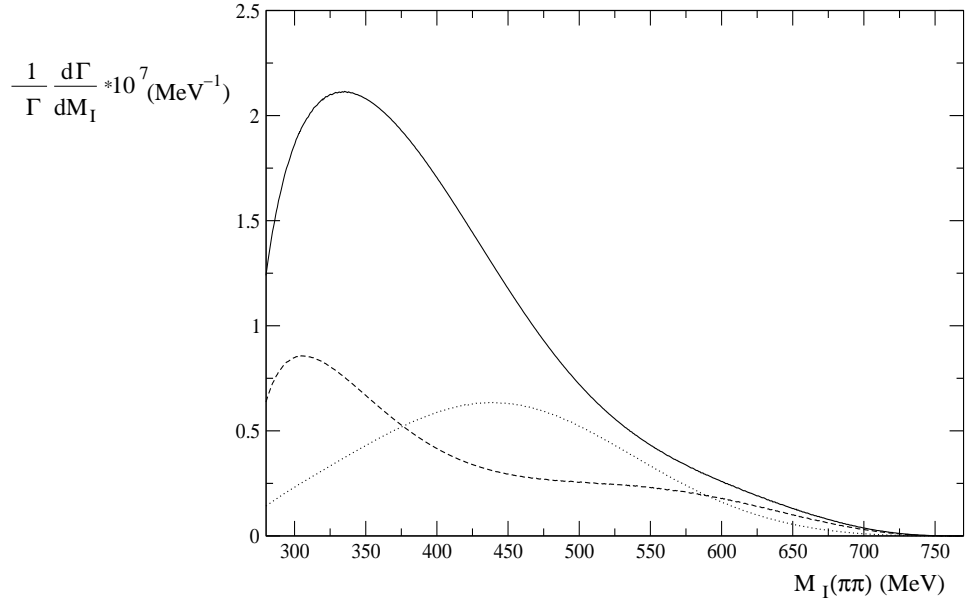,width=0.96\textwidth,silent=}}}
\caption{$dB(\rho \rightarrow \pi^{0} \pi^{0} \gamma)/dM_{I}$ as a function of
the invariant mass of the two pions. Dashed line: sequential contribution;
dotted line: loop contribution; solid line: sum of both. The sequential and loop contributions
interfere constructively.} 
\label{plot}
\end{figure}

We show the total decay widths obtained in table \ref{res}.  There we can see,
in agreement with \cite{Bramon:2001un}, 
that the dominant contribution is the one corresponding to the sequential
mechanism in all cases except for the \ropi decay, where the loop contribution
is comparable. As we can see in figure \ref{plot} the interference between these
two contribution is constructive. It is worth noting that the final shape of the
mass distribution with the sum of the sequential and loop contributions is
rather different from the one obtained with either of the two mechanisms.
Experimental information on this observable would thus be most welcome. 
The loop contribution for the \omepi decay 
was estimated small in
\cite{Bramon:2001un} using qualitative arguments. Here we corroborate this claim
 performing the actual calculation and find it also small in the case of the
\roeta and \omeeta decays, due to the relatively high mass of the kaons.
 The $\rho$-$\omega$ mixing effects are 
negligible in the  \ropi decay, but relevant in the rest of the decays.
 Although the
mixing contribution is by itself small, it has important interferences with the
sequential contribution, and in addition modifies the resonance propagators
involved, as was already mentioned in section 3.1.  

\begin{center}
\begin{table}
\begin{tabular}{|l|l|l|l|l|}
\hline
BR &  \ropi &  \roeta & \omepi & \omeeta \\
\hline
sequential & $ 1.5\cdot 10^{-5}$ & $6.6\cdot 10^{-10}$ & $4.3\cdot 10^{-5}$ &
$3.0\cdot 10^{-7}$ \\
loops & $1.5\cdot 10^{-5}$ & $5.4 \cdot 10^{-11}$ & $4.3 \cdot 10^{-7}$ &
$2.2\cdot 10^{-9}$ \\
sequential +& & & & \\$\rho$-$\omega$ mixing & not evaluated & $7.5\cdot 10^{-10}$
& $4.8\cdot 10^{-5}$
& $3.4\cdot 10^{-7}$ \\
\hline
Total & $4.2 \cdot 10^{-5}$ & $7.5\cdot 10^{-10}$ & $4.7\cdot 10^{-5}$ &
$3.3\cdot 10^{-7}$ \\ 
\hline
\end{tabular}
\caption{Branching ratios due to the different contributions to the $V
\rightarrow P^{0}P'^{0}\gamma$ decays considered.}
\end{table}
\label{res}
\end{center}

The result obtained here for the \ropi branching ratio is in good agreement with
the recent SND collaboration measurement \cite{Achasov:2000ym}: $B(\rho \rightarrow
\pi^{0} \pi^{0} \gamma) = (4.8^{+3.4}_{-1.8}\pm 0.2)\times 10^{-5}$. The same
collaboration obtained for the \omepi decay a branching ratio $B(\omega
\rightarrow \pi^{0} \pi^{0} \gamma) = (7.8\pm 2.7 \pm 2.0) \times 10^{-5}$, thus
 confirming the previous and more accurate measurement of the GAMS
collaboration \cite{Alde}, $B(\omega
\rightarrow \pi^{0} \pi^{0} \gamma) = (7.2\pm 2.5) \times 10^{-5}$. Our
predicted branching ratio for this decay is within the error bars of these experimental
 values, as we can see in table \ref{res}.  We should also mention that
 although we do not give errors in our numbers, an estimate  of $20\%$
 theoretical error is realistic in view of the accuracy of the chiral 
 Lagrangians
 used in eq. (\ref{lagr}) to provide the radiative decays of the vector mesons
 \cite{otro}. For the case of the \roeta and \omeeta decays there are not
 experimental data for the branching ratios. 
 
 Finally, in table~\ref{comparebr} we
 compare our results with other analyses and also with the experiment.
 The theoretical approaches followed in the literature are quite varied. In
 ref.~\cite{renard} current algebra, hard pions and Ward identities were used.
 In \cite{Fajfer:cd} an approach with low energy effective Lagrangians with
 gauged Wess-Zumino terms was followed. A different procedure was followed in
 \cite{Prades:1993ys} using chiral Lagrangians and the extended
 Nambu-Jona-Lasinio model to fix the couplings of the resonance contribution
 (the results given in table \ref{comparebr} are the largest ones in the
 intervals given in \cite{Prades:1993ys}, which are still low compared with
 experiment). In \cite{Bramon:1992kr} only the sequential VMD mechanisms were
 used and the results were improved in \cite{Bramon:1992ki} for the \ropi decay
 including the one loop $\chi PT$ contribution. This latter point is 
 further improved in \cite{Bramon:2001un}, where a more realistic $\pi \pi$
 isoscalar amplitude is used. In \cite{Marco:1999df} only the loop contributions
 were evaluated, and here the sequential VMD mechanisms are considered in
 addition. Finally, in \cite{Guetta:2001ra}, where only the \omepi decay is
 evaluated, the sequential VMD mechanisms are considered including the $\omega
 -\rho$ mixing, but only the sequential mechanism is used for the $\rho$ decay
 in this latter term. The order of magnitude in the different approaches is
 similar, with the exception of the results in \cite{Prades:1993ys} which seem
 abnormally low. Yet, the 
 rates obtained in \cite{Bramon:2001un} and in the present work match better with the
 experimental \ropi decay width. Table \ref{comparebr} shows also the evolution
 of the approaches and the results with the time, and how the original VMD
 mechanisms suggested in \cite{singer} have survived, while the advent of $\chi
 PT$ and its unitary extensions have brought the mechanisms needed to obtain a
 satisfactory result for the \ropi decay, as seen in \cite{Bramon:2001un} and the
 present work. With respect to this last reference our approach adds the novel
 thing of using directly the $\pi \pi$ amplitudes from $U\chi PT$, while in
 \cite{Bramon:2001un} a phenomenological model for the $T$ matrix accounting
 explicitly for the $\sigma$ and $f_{0}(980)$ mesons was used. In our approach
 both mesons are dynamically generated from the multiple scattering of pions and
 kaons driven by the dynamics of the lowest order chiral Lagrangian.

\begin{center}
\footnotesize{\begin{table}
\begin{tabular}{|l|l|l|l|l|}
\hline
Work &  \ropi &  \roeta & \omepi & \omeeta \\
\hline
\cite{Fajfer:cd} & $2.9\times 10^{-5}$ & $4.0 \times 10^{-6}$ &
$8.2\times 10^{-5}$ & $6.3\times 10^{-6}$ \\

\cite{renard} & $1.1\times 10^{-5}$ & $---$ & $2.7\times 10^{-5}$ & $---$ \\

\cite{Prades:1993ys} & $4.7\times 10^{-6}$ & $2.0\times
10^{-10}$ & $1.4\times 10^{-5}$ & $8.3\times 10^{-8}$ \\

\cite{Bramon:1992kr} & $1.1\times 10^{-5}$ & $4\times 10^{-10}$
& $2.8\times 10^{-5}$ & $1.6\times 10^{-7}$ \\

\cite{Bramon:1992ki} & $2.6\times 10^{-5}$ & $4\times 10^{-10}$
& $2.8\times 10^{-5}$ & $1.6\times 10^{-7}$ \\

\cite{Marco:1999df} & $1.4\times 10^{-5}$ & $---$ & $---$ & $---$ \\

\cite{Guetta:2001ra} & $---$ & $---$ & $(4.6\pm 1.1)\times
10^{-5}$ & $---$ \\

\cite{Bramon:2001un} & $3.8\times 10^{-5}$ & $---$ & $(4.5\pm
1.1)\times 10^{-5}$ & $---$ \\
\hline
This work & $4.2\times 10^{-5}$ & $7.5\times 10^{-10}$ & $4.7\times 10^{-5}$ &
$3.3\times 10^{-7}$ \\
\hline
Experiment & $(4.8^{3.4}_{-1.8}\pm 0.2)$  & &
$(7.8\pm 2.7 \pm 2.0) \times 10^{-5}$ \cite{Achasov:2000ym} & \\
 & $\times 10^{-5}$ \cite{Achasov:2000ym} & & $(7.2\pm 2.5) \times 10^{-5}$ \cite{Alde} & \\
 \hline
\end{tabular}
\caption{Branching ratios of the different $V
\rightarrow P^{0}P'^{0}\gamma$ decays in the literature.}
\label{comparebr}
\end{table}}
\end{center}

\begin{figure} 
\centerline{\protect\hbox{
\psfig{file=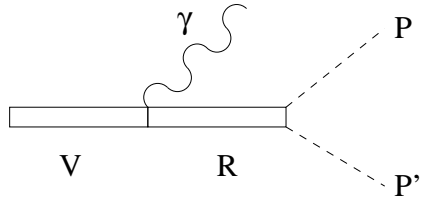,width=0.4\textwidth,silent=}}}
\caption{Feynman diagram corresponding to the $a_{2}$, $f_{2}$ meson exchange
contribution. $V$ is a vector meson, $R$ is a $f_{2}$ or $a_{2}$ resonance and P, P' can be
either $\pi^{0}$ or $\eta$ mesons.} 
\label{a2f2}
\end{figure}

The unitary approach followed here for the meson meson interaction leads to the
$f_{0}(980)$ and $a_{0}(980)$ resonances for the $\pi^{0} \pi^{0}$ and 
$\pi^{0} \eta$ final states, respectively, without having to introduce them
explicitly. One may wonder what would be the contribution of the $f_{1}(1285)$,
$a_{1}(1260)$, $f_{2}(1270)$ and $a_{2}(1320)$ resonances as intermediate
states. Avoiding the discussion whether they could or could not be generated
dynamically as their $f_{0}$, $a_{0}$ partners \cite{Dobado:2001rv}, we can deal with those
mechanisms by considering tree level Feynman diagrams which rely upon empirical
couplings of these resonances to vector mesons, pseudoscalars and photons. One
of the possible mechanisms would then be the one of fig.~\ref{feyn} where $V'$
is substituted by any of those resonances (with zero charge). However, the $VRP$
vertex, with $V$ any neutral vector meson, $P=\pi^{0}, \eta$
and $R=f_{1}, f_{2}, a_{1},a_{2}$ (with zero charge), is not allowed by charge
conjugation \cite{LopezCastro:2001qa}. Analogously one can see in
the explicit Lagrangians involving the $a_{1}$, $f_{1}$ resonances that such
terms vanish \cite{Prades:1993ys}. Thus, we are left with the diagrams where a
photon is produced in the first place, see fig~\ref{a2f2}. The topology of this
diagram is
actually the same one as that in the $f_{0}$, $a_{0}$ production considered so
far. Next we see that if $R$ is the $f_{1}$ or $a_{1}$ resonance the $RPP'$
vertex with $P, P'=\pi^{0}, \eta$ is forbidden by parity reasons, because the
decay must proceed in p-wave and $f_{1}$, $a_{1}$ have positive parity.

Finally we are left with the mechanism of fig.~\ref{a2f2} with the $f_{2}$ or
$a_{2}$ resonances. In order to estimate the contribution of these mechanisms to
the decay processes studied here, or to the radiative $\phi$ decay of 
\cite{Marco:1999df}, we rely upon the results obtained in \cite{Oller:1997yg} in
the study of the $\gamma \gamma\rightarrow \pi^{0} \pi^{0}$, $\pi^{0} \eta$,
with obvious similarity to the $V\rightarrow \gamma PP'$ process studied here. In this
work the contribution of the $f_{2}$ and $a_{2}$ resonances was considered and
found to be very important in the regions around the corresponding resonance
poles. However, as the calculation provides, or one can see explicitly in figs.
7 and 10 of \cite{Oller:1997yg}, the extrapolation of the resonance contribution
down to energies below the $\rho$ and $\omega$ mass, which we have in the decays
studied here, is negligible. At these energies the effect of the $f_{0}$,
$a_{0}$ resonances which are closer in energy are relatively more important, and
even then they do not play a significant role in the $\rho$ and $\omega$
radiative decay. 

Although we have not studied here the radiative $\phi$ decay, it is worth taking
advantage of the former discussion to estimate the effects of the $f_{2}$, 
$a_{2}$ intermediate states in this decay. Once again the figures of
ref.~\cite{Oller:1997yg} mentioned above are illustrative. Fig.~10 shows that 
for $\pi^{0} \eta$ production the $a_{2}$ contribution would be a small fraction
of the $a_{0}$ contribution at energies below 1020 MeV found in this decay. The
case of the $f_{2}$ resonance is more subtle because as one can see in fig.~7 of
\cite{Oller:1997yg} the $f_{0}$ resonance shows up only weakly in the $\gamma
\gamma \rightarrow \pi^{0} \pi^{0}$ reaction and the
background of the $a_{2}$ resonance at energies around 1 GeV does not seem
negligible. However, the small signal of the $f_{0}$ resonance in this reaction
is due, as noted in \cite{Oller:1997yg}, to a strong cancellation between the
$f_{0}$ contribution and the one with an $\omega$ in the intermediate state in
the $\gamma \gamma \rightarrow \pi^{0} \pi^{0}$ process. However, while the
$\gamma \omega \pi^{0}$ vertex is sizeable, the $\phi \omega \pi^{0}$ vertex is
not allowed by isospin conservation, and hence the cancellation does not appear
in the radiative decay of the $\phi$, making the $f_{0}$ production sizeable as
clearly seen in the experimental data 
\cite{Aulchenko:1998xy,Akhmetshin:1999di,Achasov:2000ym} which are dominated by
the $\phi\rightarrow f_{0} \gamma$ decay. In the absence of that cancellation
the $f_{2}$ contribution to the radiative decay of the $\phi$ into $\pi^{0}
\pi^{0} \gamma$ is also very small compared to the dominant $\phi \rightarrow
f_{0} \gamma$ contribution.

\section{Conclusions}

We have studied the radiative decays of the $\rho$ and $\omega$ mesons into two
neutral mesons including the mechanisms of sequential vector meson decay,
$\rho$-$\omega$ mixing and chiral loops. In the case of the $\omega$ decays we
find, confirming previous findings, that the sequential mechanism is dominant.
We also find that the loop contribution is very small for the \roeta case. The
$\rho$-$\omega$ mixing was found non negligible for the $\rho \rightarrow
\pi^{0} \eta \gamma$, \omepi and \omeeta decays. The loop contribution is very
important in the case of the $\rho \rightarrow \pi^{0} \pi^{0} \gamma$ and the 
branching ratio obtained in this case, with the sum of the sequential and loop
mechanisms, is about three times larger than with either mechanism alone,
leading to  values compatible with present experimental measurements. The shape
 of
 the $\pi \pi$ mass distribution is also found rather different to that with either
of the mechanisms alone. We have also estimated the effects of the $f_{2}$,
$a_{2}$ intermediate resonances concluding that they are negligible for the
decays studied here. In the case of the \omepi decay our predicted branching
ratio is also in agreement with the experimental values. The results obtained
for the \ropi decay provide a further consistency check of the $U\chi PT$
approach to meson-meson interaction and its underlying interpretation of the
nature of the scalar mesons. Further measurements on invariant mass
distributions would provide extra tests of these ideas, allowing us to gain
further insight on the controversial nature of the scalar mesons.

\subsection*{Acknowledgments}

One of us, S.H. wishes to acknowledge the hospitality of the University
of Valencia where this work was done and financial support from the Fundaci\'on
BBV. We are also grateful to A. Bramon for useful discussions. J.E.P. 
acknowledges support from the Ministerio de Educaci\'on y Cultura.
This work is also
partly supported by DGICYT contract number BFM2000-1326, E.U. EURODAPHNE
network
contract no. ERBFMRX-CT98-0169 and
by the Grants-in-Aid for Scientific Research of the Japan Ministry of
Education, Culture, Sports, Science and  Technology (No. 11440073 and No.
11694082).

\end{document}